\documentclass{elsarticle}

\usepackage{graphicx}
\usepackage{epstopdf}
\usepackage{amssymb}
\usepackage{caption,booktabs}
\usepackage{subcaption}
\usepackage{lineno}
\usepackage{color}
\usepackage{soul}

\biboptions{square}

\begin{document}

\begin{frontmatter}



\title{Charge separation
in organic photovoltaic cells}


\author[label1]{Paraskevas Giazitzidis}
\author[label1]{Panos Argyrakis}
\author[label2]{Juan Bisquert}
\author[label3]{Vyacheslav S. Vikhrenko}

\address[label1]{Department of Physics, University of Thessaloniki, Thessaloniki, Greece}
\address[label2]{Photovoltaics and Optoelectronic Devices Group, Departament de Física, Universitat Jaume I, 12071 Castelló, Spain}
\address[label3]{Belarusian State Technological University, Minsk, Belarus}

\begin{abstract}
We consider a simple model for the geminate electron-hole separation process in organic photovoltaic cells, in order to illustrate the influence
of dimensionality of conducting channels on the efficiency of the process. The Miller-Abrahams expression for the transition rates between
nearest neighbor sites was used for simulating random walks of the electron in the Coulomb field of the hole. The non-equilibrium kinetic Monte
Carlo simulation results qualitatively confirm the equilibrium estimations, although quantitatively the efficiency of the higher dimensional
systems is less pronounced. The lifetime of the electron prior to recombination is approximately equal to the lifetime prior to dissociation. Their values indicate
that electrons perform long stochastic walks before they are captured by the collector or recombined. The non-equilibrium free energy
considerably differs from the equilibrium one. The efficiency of the separation process decreases with increasing the distance to the collector,
and this decrease is considerably less pronounced for the three dimensional system. The simulation results are in good agreement with the
extension of the continuum Onsager theory that accounts for the finite recombination rate at nonzero reaction radius and non-exponential
kinetics of the charge separation process.
\end{abstract}


\end{frontmatter}

\section{Introduction}
Photovoltaic cells utilizing organic semiconductors have attracted much attention due to their promising electronic properties, low cost, thin
film flexibility, and high functionality. Although their present efficiency is not high enough, widespread interest in both the academic and,
increasingly, the commercial communities promises fast progress in this direction \cite{Klarke2010,Deibel2010}. Many attempts have been
undertaken to understand the dissociation and recombination processes starting with the theoretical works of Onsager
\cite{Onsager1934,Onsager1938}, Frenkel  \cite{Frenkel1938} and Eigen  \cite{Eigen1954}, where models of these processes were suggested and
investigated. Later, these models were refined with using proper boundary conditions for more adequate description of different stages of
recombination/dissociation \cite{Noolandi1978,Noolandi1979,Sano1979,Wojcik2009}. In these models it is assumed that continuum
phenomenological diffusion equations are applicable on molecular space and time scales. On the other hand, discrete models were developed and
their computer simulations were performed \cite{Scher1984,Albrecht1995,Peumans2004,Deibel2009,wojcik2010}.

The models of charge separation were widely used to find routes for increasing the efficiency of the organic solar cells whose internal
structure is characterized by a variety of characteristics, e. g. the charge transfer routs can be of different space dimensionality. For
example, discotic liquid-crystal porphyrins conduct almost exclusively along one-dimensional backbones, most $\pi$-conjugated polymers have
quasi-two-dimensional character, while C$_{60}$ and their derivatives are truly three-dimensional organic semiconductors \cite{Gregg2011}.

The effect of space dimensionality on recombination/dissociation was discussed earlier \cite{barzykin1993}.
It was shown that in the absence of interaction potential between two dissociating particles the escape probability (the probability that the particles can go away to infinity) is zero if the space dimension is smaller than or equal to two and increases with increasing space dimensionality.

Recently the entropy contribution with respect to the dimensionality of the organic semiconductor into charge separation after sunlight exciton
production has been extensively discussed \cite{Klarke2010,Gregg2011}. The equilibrium free energy of the electron-hole pair was used for
estimating the charge separation efficiency in systems of different dimensionality. It was shown that the efficiency of three dimensional (3D)
systems can be up to four orders of magnitude higher, as compared to an one dimensional (1D) system.

However, the equilibrium consideration does not take into account important features of the charge separation process as it was mentioned in
Ref. \cite{Gregg2011} and is in more detail discussed below. Thus, we suggest a simple non-equilibrium model of geminate
recombination/dissociation of electron-hole pairs and quantitatively investigate the electron yield on the collector depending on the system
dimensionality with accounting of interparticle Coulomb interaction. The main focus of our work is the influence of the space dimensionality on
the electron yield in zero electric field. The simulation results are compared with the extension of the Onsager continuum model
\cite{Sano1979,Wojcik2009}. Characteristic times of the processes are considered as well.

\section{Model description}
To make the model as simple as possible we consider an electron that moves in a regular lattice and in the Coulomb field of the immobile
hole \cite{Klarke2010,Gregg2011}. Then, the electron can be considered as moving in the external Coulomb field of the hole and in the approximation
of uniformly distributed lattice sites, its equilibrium distribution function can be written as
\begin{equation}
 f(r)=Q^{-1}n_r \exp[-\beta u(r)]=Q^{-1} \exp[-\beta [u(r)-k_B T S(r)]] ,
 \label{eq1}
\end{equation}
where $Q$ is a normalization constant; $n_r=1,2\pi r, 4\pi r^2 $ for one, two and three dimensional systems, respectively, and it determines the
density of the number of lattice sites as a function of $r$ that can be occupied by the electron with equal probability; $r$ is the
electron-hole distance; $T$ is the absolute temperature; $k_B$ Boltzmann constant; $\beta=1/(k_BT)$ is the inverse temperature. The energy of
Coulomb interaction for the electron on a cubic lattice with the lattice parameter $a$ is inversely proportional to electron-hole distance
\begin{equation}
 u(r)= \frac{-e_0}{\sqrt{i^2+j^2+k^2}} , e_0=\frac{e^2}{4\pi \varepsilon_0 \varepsilon a} ,
 \label{eq2}
 \end{equation}
 where $\varepsilon$ is the medium dielectric constant; $e$ electron charge; $\varepsilon_0$ electric constant; $i,j,k$ are integers that
 determine lattice sites positions. For 1D and 2D systems the expression for the interaction energy has to be modified correspondingly.
The second part of Eq.(\ref{eq1}) is rewritten in such a way that the configuration entropy $S(r) = \ln n_r$ appears in a natural way.
The expression in the square brackets determines the Helmholtz free energy
\begin{equation}
\Delta G(r) = u(r)-k_B T S(r) ,
 \label{eq9}
\end{equation}
 that cumulatively takes into account the equal probability of occupying a number of equivalent lattice sites (with the same distance $r$) and the attractive
electron-hole interaction that makes smaller $r$ distances much more preferable. Of course, this expression for $\Delta G$ is the same
as given in \cite{Klarke2010,Gregg2011}.
The distribution function is normalized to unity and the normalization constant
\begin{equation}
Q=\sum_{i,j,k}\exp(-u(i,j,k)/k_BT) ,
\label{eq3}
\end{equation}
where the sum runs over all sites between the source and the collector.

We used kinetic Monte Carlo method with the Metropolis algorithm \cite{Metropolis1953} to perform simulations of the equilibrium distribution
functions. The parameters were taken from paper \cite{Gregg2011}: $\varepsilon=4$, $a=1$nm, $T=300$K. Then, the characteristic energy
$e_0/k_BT\cong13.9$ and the normalization constants are $Q\cong(1.0882; 4.4448; 6.8888)\cdot 10^6$ for 1D, 2D and 3D cases, respectively. In the
equilibrium simulation the electron can jump from the initial site to any other site inside of the collector $(i^2+j^2+k^2 \leqslant N^2+N)$
except zero site ($i^2+j^2+k^2 =0$ where the hole is situated). The lattice sites are prescribed to the distance $r$ if (in 3D case with the
corresponding changes in 2D and 1D cases) $r^2 - r < i^2+j^2+k^2 \leqslant r^2 + r$, $i,j,k,r$ and $N$ are integers and the collector is on the
distance $(N+0.5)a$ from the hole. The configuration entropy in Monte Carlo simulation is taken into account indirectly through the
interrogation of the lattice sites.

The electron energy difference $\Delta u$ between the destination and the initial state was used for calculating the probability of the electron
transition between these two sites. The transition probability was taken equal to $\exp(-\Delta u/k_BT)$ if $\Delta u>0$ or 1 if $\Delta u\leqslant0$.
The free energy was calculated from the probabilities for the electron to occupy the lattice sites in accordance with Eq.\ref{eq1}
\begin{equation}
\Delta G(r)= -k_BT[ \ln f(r) + \ln Q - \ln n_r] .
\label{eq4}
\end{equation}

Definitely, in a true geminate recombination/dissociation process the electron cannot be in equilibrium with the hole. However, if the
recombination rate is very small and can be neglected then the equilibrium model can be used for crude estimation of the electron spatial
distribution around the hole.

\section{Results and discussion}
\subsection{Equilibrium results}
The results are shown in Fig.\ref{fig1a} for $N=30$ (for the sake of uniformity in distribution of the lattice sites over distance $r$, the
distance to the collector was taken equal to $30.5a$) and they confirm the equilibrium distribution function (\ref{eq1}) and the conclusion
of papers \cite{Klarke2010,Gregg2011} about higher efficiency of 3D systems. Small fluctuations of the Monte Carlo simulation results around
the analytical curves are explained by nonhomogeneous distribution of lattice sites over $r$ distances.
The efficiency of charge separation in the equilibrium consideration was defined\cite{Klarke2010,Gregg2011} as the ratio of the probabilities
for the electron to be on a certain distance $R$ from the hole and at its ground state that for Boltzmann statistics is proportional to
exp$(-\Delta E/k_BT)$, where $\Delta E$ is the free energy difference between these electron positions.

In Fig.\ref{fig1b} the results for the systems with the distance of $500a$ from the source to the collector are shown.
It is evident that the efficiency of the systems estimated from the equilibrium free energy strongly depends on the distance between the source and the collector. The longer is this
distance, the larger is the difference by many orders of magnitude in the efficiency of the systems of different dimensionality. Moreover, for
2D and 3D systems the longer is the distance from the source to the collector, the larger are the absolute values of the charge separation
efficiency. This is counterintuitive to the real situation.

\subsection{Nonequilibrium results for the electron yield at the collector}
However, the equilibrium results do not take into account the finite lifetime of the electron before it arrives at the collector or recombines
with the hole. Thus, non-equilibrium simulations are in order. Such simulations have been published earlier
\cite{Albrecht1995,Peumans2004,Deibel2009}, however they were mainly focused on the electric field dependence of the electron yield. Analytical
results are available as well \cite{barzykin1993} and they indicate that the larger the dimensionality is the
larger survival probability for the electron exists at all distances. Moreover, $D$=2 is the critical dimensionality below which the
electron-hole pair cannot dissociate (in the sense that the electron-hole distance tends to infinity) at non-zero recombination rate in the
absence of the external field.

We have chosen as starting point of the electron the site close to the hole (on 1nm distance) and $10^6 - 10^7$ electron trajectories were
considered for each particular simulation to ensure good statistics. The simulations were performed for the system described above. The hole was
fixed at the coordinate origin and the electron collector was taken at the distance of $30.5a$. Initially the electron was placed on the nearest
neighbor to the origin's site. The transition rate for the electron to recombine from this initial site is designated as $w_{10}$. The
transition rate of the electron to its other nearest neighbor site was calculated in accordance to Miller - Abrahams \cite{Miller1960}
expression
\begin{equation}
w_{n\rightarrow n+1}=w_0\exp[-(u_{n+1}-u_n+\vert u_{n+1}-u_n \vert)/2k_BT] ,
\label{eq5}
\end{equation}
\begin{equation}
w_0 = v_0\exp(-2a/\alpha) ,
\label{eq6}
\end{equation}
where $u_n$ and $u_{n+1}$ are the electron energies at the initial and destination sites, respectively; $\alpha$ is the electron localization
parameter; $w_0$ the hopping attempt frequency; $w_0$ determines the time scale of the charge separation process and the simulations were
performed at $w_0=1$; then all the other rates are given in units of $w_0$. Thus, our non-equilibrium model corresponds to the discrete version of the Onsager recombination model \cite{Onsager1938}
with accounting of finite time and non-zero reaction radius of geminate recombination according to \cite{Noolandi1978,Noolandi1979,Sano1979} at
zero electric field.

Although we consider the hole fixed at the origin, the results can be used for situations when both charges are moving with the proper
definition of the diffusion constants \cite{Noolandi1979,Wojcik2009}. On the other hand, the simulation results are valid for the system with
separated electron subspace by reflecting boundary conditions for the plane passing through the immovable hole. This means that if the electron
occupies site $i=0 , \vert j \vert + \vert k \vert \neq 0 $ and the trail is $i=-1$ then it has to be taken $i=1$.

We can consider three possibilities: $(a)$ the electron of the just created exciton can immediately or after some waiting time recombine; $(b)$
the electron can recombine after random walks over the lattice at the influence of a quite strong Coulomb electron-hole interaction; $(c)$ the
electron can reach the collector (dissociation). The characteristic time of process $(a)$ is estimated as $w_{10}^{-1}$ . However, the total
recombination rate is determined by both $(a)$ and $(b)$ processes. The distribution of the recombination and the electron arrival at the
collector times can easily be extracted from Monte Carlo simulation results.

The simulation results are shown in the Table and, for visibility, in Fig.\ref{fig2a}. The recombination rates (the electron transition
to the origin in process $(a)$) are shown in the first column. For comparison, the dimensionless transition rate for an electron from
the nearest to the origin site to its second neighbor in 1D system is equal to $w_{12}=\exp( -13.9 / 2)\cong 0.00096$, while in 2D and 3D
systems it is considerably larger, $w_{13}= \exp( -13.9 (1 - 2^{-1/2})) \cong 0.017$; the transition rates grow quickly with increasing
the electron-hole distance. The next three columns in the Table contain the electron yields at the collector (the ratios of the number of electrons reached the collector to the total number of electrons for each particular value of $w_{10}$) in 1D, 2D and 3D systems,
respectively. The three last columns contain the ratios of the electron yields for the systems shown in the column headers. These ratios
are almost constant in wide range of the recombination rates up to the point where the yield achieves values larger than approximately $0.1$.
Some fluctuations in the last column are due to poor statistics of 1D systems at low electron yields.
\begin{center}
\begin{table}
  \captionof{table}{The electron yields and their ratios for systems of different dimensionality.}

 \begin{tabular}{l*{6}{c}r}
 $w_{10}$              & 1D & 2D & 3D & 3D(Eq.\ref{eq7}) & 2D/1D  & 3D/2D & 3D/1D \\
 \hline
 0.01           & $7.00\cdot 10^{-6}$ & $2.23\cdot 10^{-4}$ & $3.85\cdot 10^{-3}$ & $3.85\cdot 10^{-3}$ & 31.8 & 17.3 & 550  \\
 0.005          & $1.27\cdot 10^{-5}$ & $4.40\cdot 10^{-4}$ & $7.65\cdot 10^{-3}$ & $7.67\cdot 10^{-3}$ & 34.6 & 17.4 & 602  \\
 0.002          & $3.45\cdot 10^{-5}$ & $1.13\cdot 10^{-3}$ & $1.89\cdot 10^{-2}$ & $1.90\cdot 10^{-2}$ & 32.8 & 16.7 & 548  \\
 0.001          & $6.77\cdot 10^{-5}$ & $2.25\cdot 10^{-3}$ & $3.70\cdot 10^{-2}$ & $3.72\cdot 10^{-2}$ & 33.2 & 16.4 & 547  \\
 0.0005         & $1.28\cdot 10^{-4}$ & $4.49\cdot 10^{-3}$ & $7.17\cdot 10^{-2}$ & $7.17\cdot 10^{-2}$ & 35.1 & 16   & 560  \\
 0.0003         & $2.22\cdot 10^{-4}$ & $7.45\cdot 10^{-3}$ & $0.114$ & $0.114$ & 33.6 & 15.3 & 514  \\
 0.0001         & $6.57\cdot 10^{-4}$ & $2.20\cdot 10^{-2}$ & $0.279$ & $0.278$ & 33.6 & 12.7 & 425  \\
 0.00005        & $1.33\cdot 10^{-3}$ & $4.32\cdot 10^{-2}$ & $0.436$ & $0.436$ & 32.5 & 10.1 & 328  \\
 0.00001        & $6.62\cdot 10^{-3}$ & $0.184$             & $0.794$ & $0.794$ & 27.8 & 4.3  & 120  \\
 \end{tabular}
\end{table}
\end{center}
Thus, it is evident that the competition of the configuration entropy contribution and Coulomb interaction gives rise to higher efficiency of
higher dimensional systems in the charge recombination/dissociation processes.

Referring to the ergodicity hypothesis it is possible to say that in equilibrium the electron is moving on the equilibrium free energy surface
and at the same time this surface determine according to Eq.(\ref{eq1}) the probability distribution of the equilibrium ensemble of electrons
over the lattice sites. Thus, we can consider as a mathematical construction the non-equilibrium free energy calculated by
\begin{equation}
\Delta G(r)=-(\ln Q + \ln f(r))k_BT , \label{eq8}
\end{equation}
where $f(r)$ is the non-equilibrium distribution function evaluated from the Monte Carlo simulation results and normalized to unity. On the
other hand, the Monte Carlo simulation procedure can be considered as reproducing the electron motion over this non-equilibrium free energy
surface.

The equilibrium and non-equilibrium free energies can be used for comparing the distribution of times spent by the electron on lattice sites
because the distribution functions themselves vary by many orders of magnitude and are not convenient for comparison. The non-equilibrium free
energy (Fig.\ref{fig3}) considerably differs from the equilibrium one. Its prominent feature is that it does not depend on the recombination
rate. This energy strongly increases near the collector that indicates significant decrease of probability distribution due to absorption of the
electrons by the collector.

\subsection{Comparison with the continuum representation}
The current simulation results can be compared with the extension \cite{Sano1979,Wojcik2009} of the continuum Onsager model that takes into
account the finite geminate recombination rate at nonzero reaction radius and the non-exponential character of the electron-hole separation
process. As we calculated the electron yield $Y(qa/a)$ on a certain distance $(qa)$ from the hole, starting from the distance $a$ it is possible
to extract the corresponding value from the expression \cite{Sano1979,Wojcik2009} for the yield at infinite distance when starting from the
distances $a$ $(Y(\infty/a))$ and $qa$ $(Y(\infty/qa))$ using the probability product rule
\begin{equation}
Y(qa/a)=Y(\infty/a)/Y(\infty/qa)=[1+\frac{pa^2}{Dr_c}(\exp(\frac{q-1}{q}\frac{r_c}{a}))]^{-1} ,
\label{eq7}
\end{equation}
where $q=N+0.5$, $p$ is a reactivity parameter, $D$ the diffusion coefficient, $r_c = e^2/(4\pi\varepsilon_0\varepsilon k_BT)$ the Onsager
radius; in our simulation $(r_c/a)=13.9$ and $D=a^2$ because the frequency factor was taken equal to 1.

\begin{figure}
\centering
\begin{subfigure}{.5\textwidth}
  \centering
  \includegraphics[width=\linewidth]{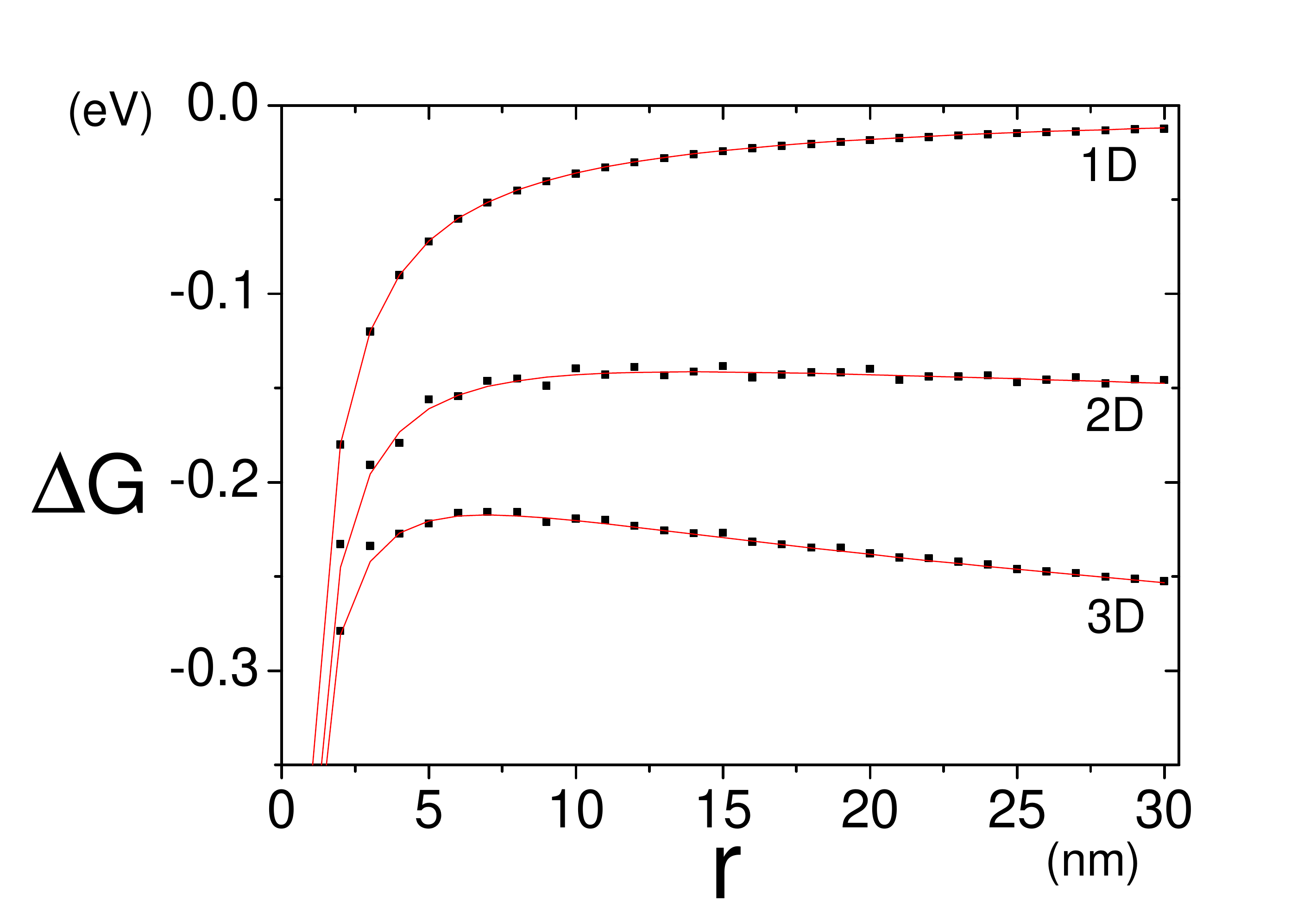}

  \caption{}
  \label{fig1a}
\end{subfigure}
\begin{subfigure}{.5\textwidth}
  \centering
  \includegraphics[width=\linewidth]{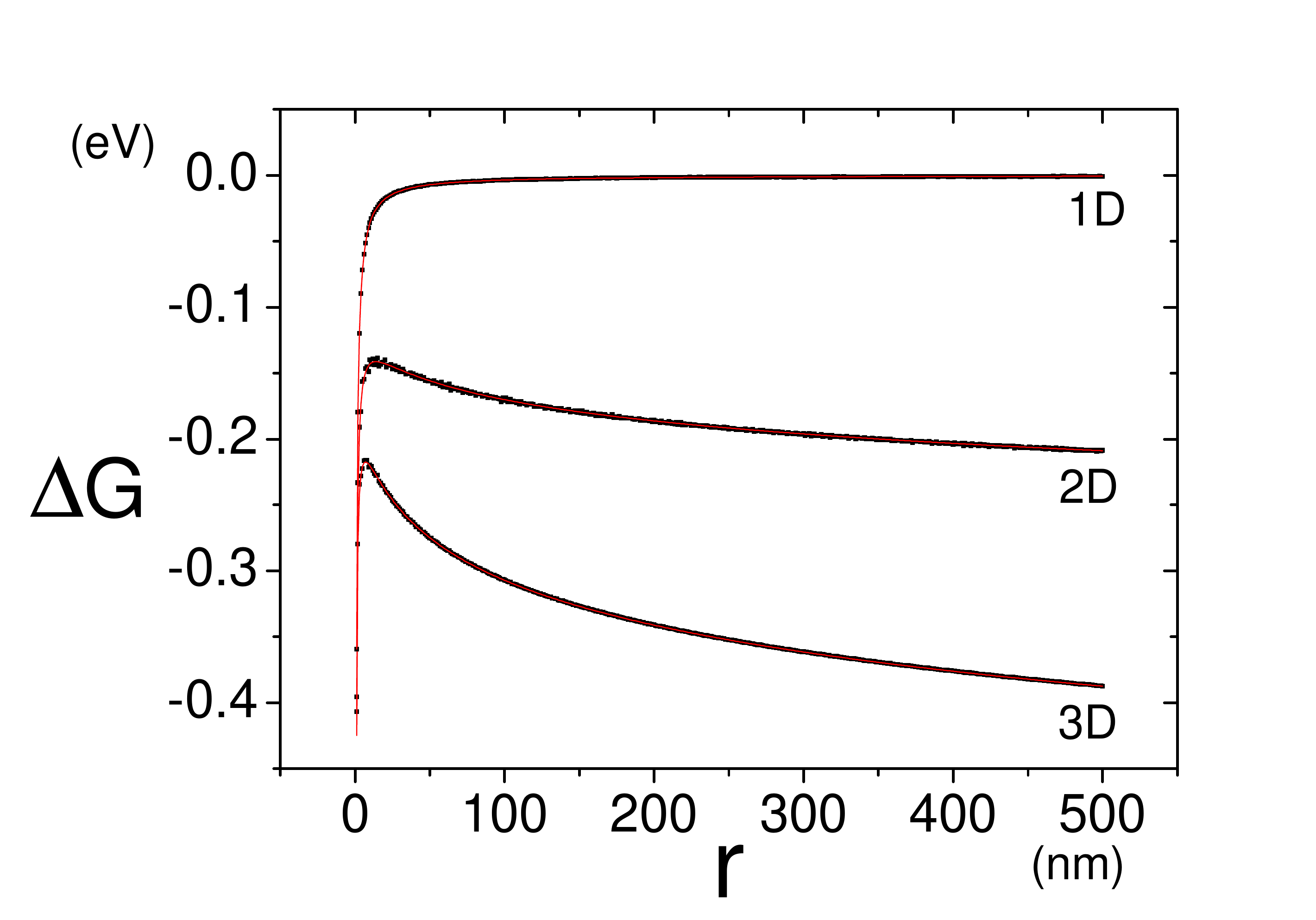}
  \caption{}
  \label{fig1b}
\end{subfigure}
\caption{(color online) The free energy versus the electron-hole distance for one, two, and three
dimensional systems of size (a) $30a$; (b) $500a$. The lines reproduce the analytic expressions of Eq.\ref{eq9}; the dots are the Monte Carlo
simulation results.} \label{fig1}
\end{figure}

The main problem in comparing our discrete model results with the continuum theory \cite{Sano1979,Wojcik2009} is the evaluation of the
reactivity parameter $p$. Comparing to the 3D case, the recombination rate as the flux through the spherical surface around the hole
\cite{Wojcik2009} and the flux from the volume $6a^3$ where recombination occurs we arrive at the estimation
$p=(3/2\pi)w_{10}a\cong0.477w_{10}a$ and then $pa^2/Dr_c\cong0.0343w_{10}$ . However, with this value of $p$ the theoretical results were
systematically higher by some $10\%$ of the simulation results for $N=30$. Thus, slightly larger values $p=0.522w_{10}a$ were used and the
calculation results are given in the fifth column of the table. As long as the second term in Eq.\ref{eq7} is around 10 or larger the yield is
inversely proportional to the recombination rate $w_{10}$. The non-equilibrium simulation results qualitatively confirm the equilibrium
estimations, although quantitatively the higher efficiency of the higher dimensional systems is less pronounced and depends on the ratio of the
recombination and the transition rates in the vicinity of the hole. In 1D system the yield is almost exactly inversely proportional to the
recombination rate, while in 2D and 3D systems the saturation effect causes deviations from such behavior at high values of the yield.

\begin{figure}
\centering
\begin{subfigure}{.5\textwidth}
  \centering
  \includegraphics[width=\linewidth]{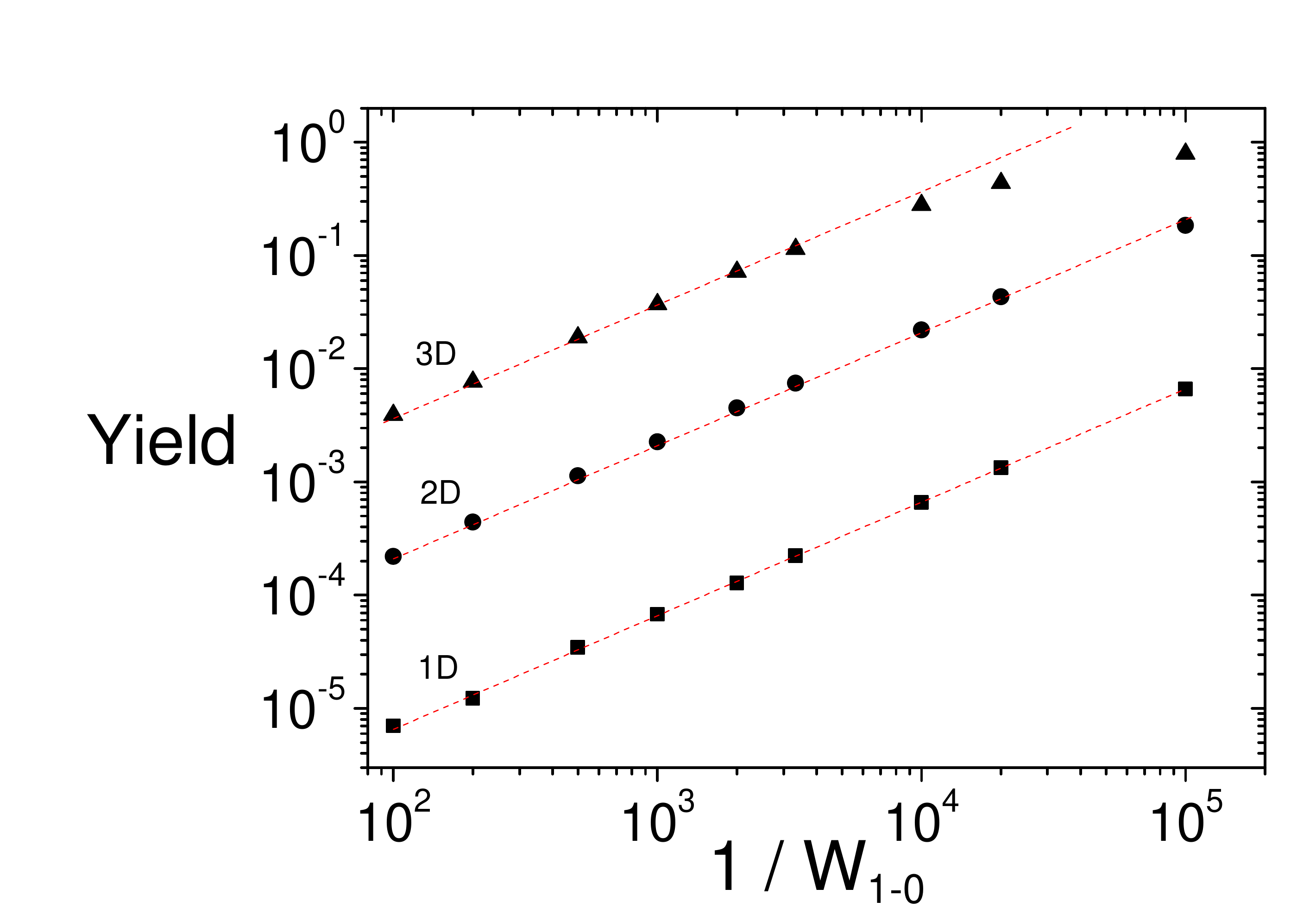}
  \caption{}
  \label{fig2a}
\end{subfigure}%
\begin{subfigure}{.5\textwidth}
  \centering
  \includegraphics[width=\linewidth]{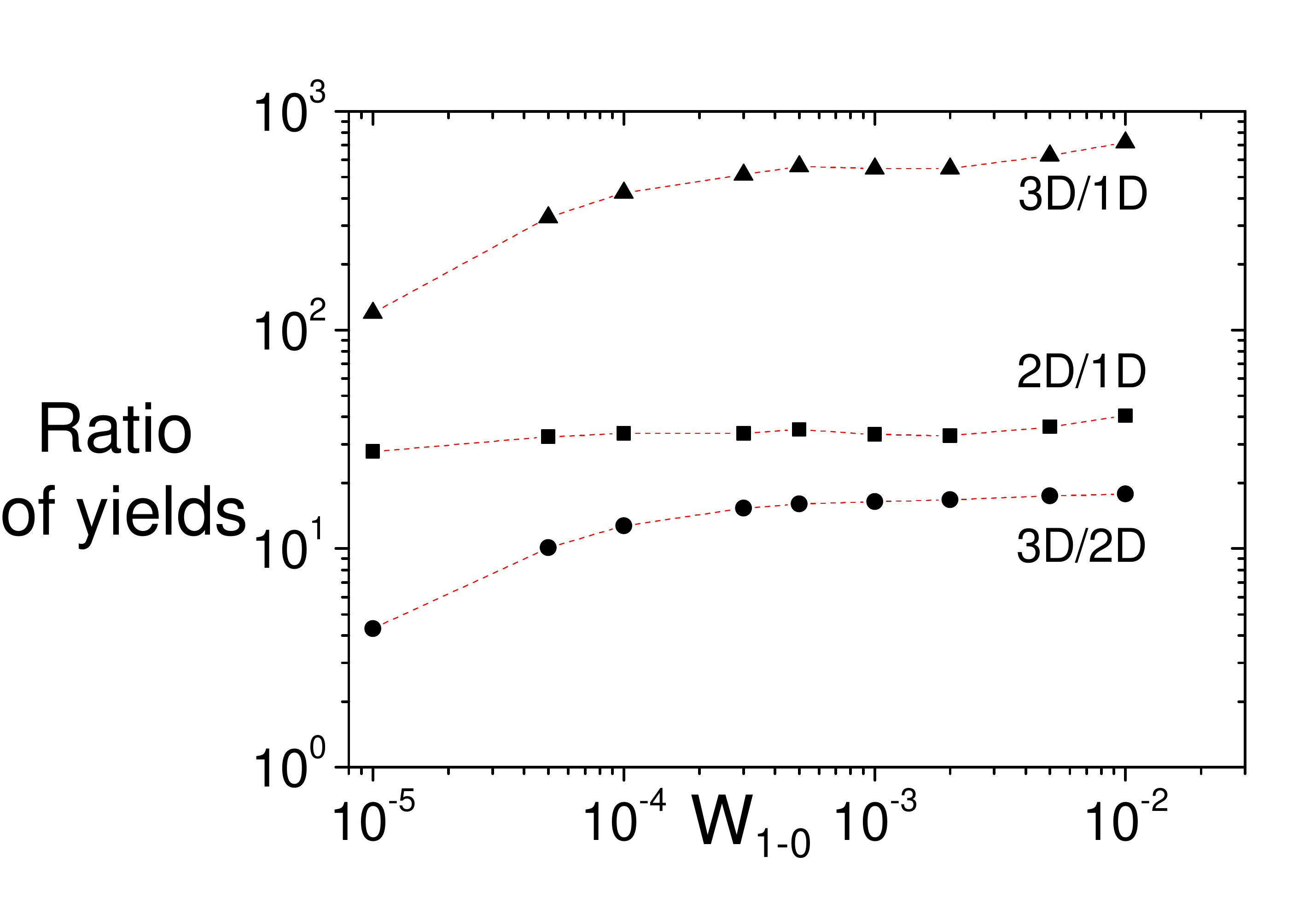}
  \caption{}
  \label{fig2b}
\end{subfigure}
\caption{(Color online) The electron yields (a) and their ratios (b) versus recombination rates for the systems of different dimensionality.
Straight dashed lines are the linear fittings, while curves are guides for the eyes.} \label{fig2}
\end{figure}

As it was mentioned above (Fig.\ref{fig1b}) the equilibrium consideration predicts an increase of the charge separation efficiency in the 2D
and 3D systems with increasing the distance to the collector, something that is counterintuitive because it does not take into account the
kinetics of the process. The non-equilibrium simulations for a larger system of $N=50$ at $w_{10}=0.0003$ have shown that the absolute values
of the electron yields are equal to $9.48\cdot10^{-5} , 4.8\cdot10^{-3},$ and $9.78\cdot10^{-2}$ for 1D, 2D, and 3D systems, respectively, and
they are smaller than the corresponding values for the system of $N=30$ by 2.27, 1.55, and 1.16 times. This means that although the absolute
values of the yields decrease with increasing distance to the collector the relative efficiency of higher dimensional systems increases. Of
course, 30 or 50 nm are too small distances as compared to that electrons have to move in modern photovoltaic devices. Nevertheless, these
distances correspond respectively to more than 2 and almost 4 Onsager radiuses and thus capture the main features of the
dissociation/recombination process.

To adjust the theoretical value of Eq.\ref{eq7} for $N=50$ with the simulation result it is necessary to take $p=0.514w_{10}a$ which is
smaller than that for $N=30$ and closer to the theoretical value $p=0.477w_{10}a$, because at larger scale the discrepancy between discrete
and continuum approaches becomes less pronounced. Moreover, separation of charges is an initial value problem and thus memory effects contribute
to diffusion on a lattice \cite{Loukas2011} even when only one particle is considered \cite{Chandrasekhar} while they are not taken into account
by the continuum diffusion equation used in the Onsager and other such models. The difference in $10\%$ only between the results of the
continuum and discrete models is surprising in view of application of continuum representations up to molecular scales.

It is interesting to note that the geminate charge recombination in the 3D heterojunction system with planar interlayer boundary can be formally
described by the system of differential equations in four dimensions \cite{wojcik2012}. On this basis the increase of the
escape probability in systems with heterojunction following from the dynamic Monte Carlo simulation \cite{Peumans2004,wojcik2010} was explained \cite{wojcik2012}. Really, if we consider for simplicity a one-dimensional heterojunction systems then
for a given electron-hole distance in $r$ lattice spacings there are $r$ energetically equivalent positions that result in exactly the same
equilibrium distribution function as for the 2D system without heterojunction. However, the situation is to some extent controversial. In
heterojunction systems the total configuration entropy of the system decreases because of braking its translational symmetry. From physical
point of view it means that in systems without heterojunction excitons can be created in any space position, while in heterojunction systems the
excitons have to be created at the interlayer boundary or the excitons have to be efficiently transported to this boundary prior to their
recombination.

The theoretical result \cite{Sano1979,Wojcik2009} can be easily used for investigating how the type of the lattice influences the electron
yield. For example, for the face centered cubic lattice for the same nearest neighbor distance $a$ the recombination volume and the diffusion
coefficient \cite{Manningbook} are equal to $6\sqrt{2}a^3$ and $2a^2$, respectively. Thus, the coefficient $(pa^2/Dr_c)$ is smaller by a factor
of $\sqrt{2}$ than the case of the simple cubic lattice, which in turn results in grater electron yield. Moreover, in this case the density of
lattice sites is larger by the $\sqrt{2}$ factor as well. For equal densities the nearest neighbor distance has to be $\sqrt[6]{2}$ times
larger, which will result in strong increase (approximately by a factor of 4 at accepted conditions) of the electron yield in the
face centered cubic lattice as compared to the simple cubic lattice. The influence of the type of the lattice on the recombination/separation
process was considered in Ref.\cite{Rackovsky1985} in another context in the model involving more complicated parameterization.

\subsection{The electron lifetimes}
The distribution of electron lifetimes before arriving at the collector is shown in Fig.\ref{fig4}. The total number of electrons was $10^7$,
however only a part of them indicated in the Table arrived at the collector. To make the results more transparent the total simulation time was
split in bins of 50 Monte Carlo steps (MCs) and the electrons were collected in each bin. A small number of electrons went quickly to the
collector while the majority arrived at the collector after long stochastic walks over the lattice. Thus, the distributions of arrival times
have maximal values at several thousand of MCs. The decaying parts of the curves were fitted by exponentially decaying functions and the
relaxation times are inversely proportional to the recombination rates as shown in the insert of Fig.\ref{fig4}; they increase with the lattice
dimensionality because the larger the dimensionality the longer electron walks are in order. The same dependence is observed for the
distribution of the recombination times as is shown in Fig.\ref{fig5}, except that this distribution decays from the very beginning. It is
worth to note that the relaxation times are approximately the same for the distributions of arrival and recombination times. For comparison, we
can calculate the portion of electrons that recombine according to process (a) as $w_{10}/(w_{10}+w_{12})$  for 1D system and as
$w_{10}/(w_{10}+w_{12}+2w_{13})$  for 2D system. For the 3D system a multiplication by 4 in the last expression must be used instead of 2. For
the system with $w_{10}=0.0003$  we get 0.238, 0.0164 and 0.0085 for 1D, 2D and 3D cases, respectively. Thus, in 1D system a significant part of
the electrons recombine immediately after the exciton creation and the details of random walks in the vicinity of the hole that depend on the
structure of the lattice strongly influence the electron yield. The recombination time of the process (a) $w_{10}^{-1}=3333$ MCs is the same for
all dimensionalities, while for (a) and (b) processes in accordance with Fig.\ref{fig5} they are 7000, 13500, and 18500 MCs for 1D, 2D and 3D
systems, correspondingly.

\begin{figure} 
\centering
\includegraphics[scale=.4]{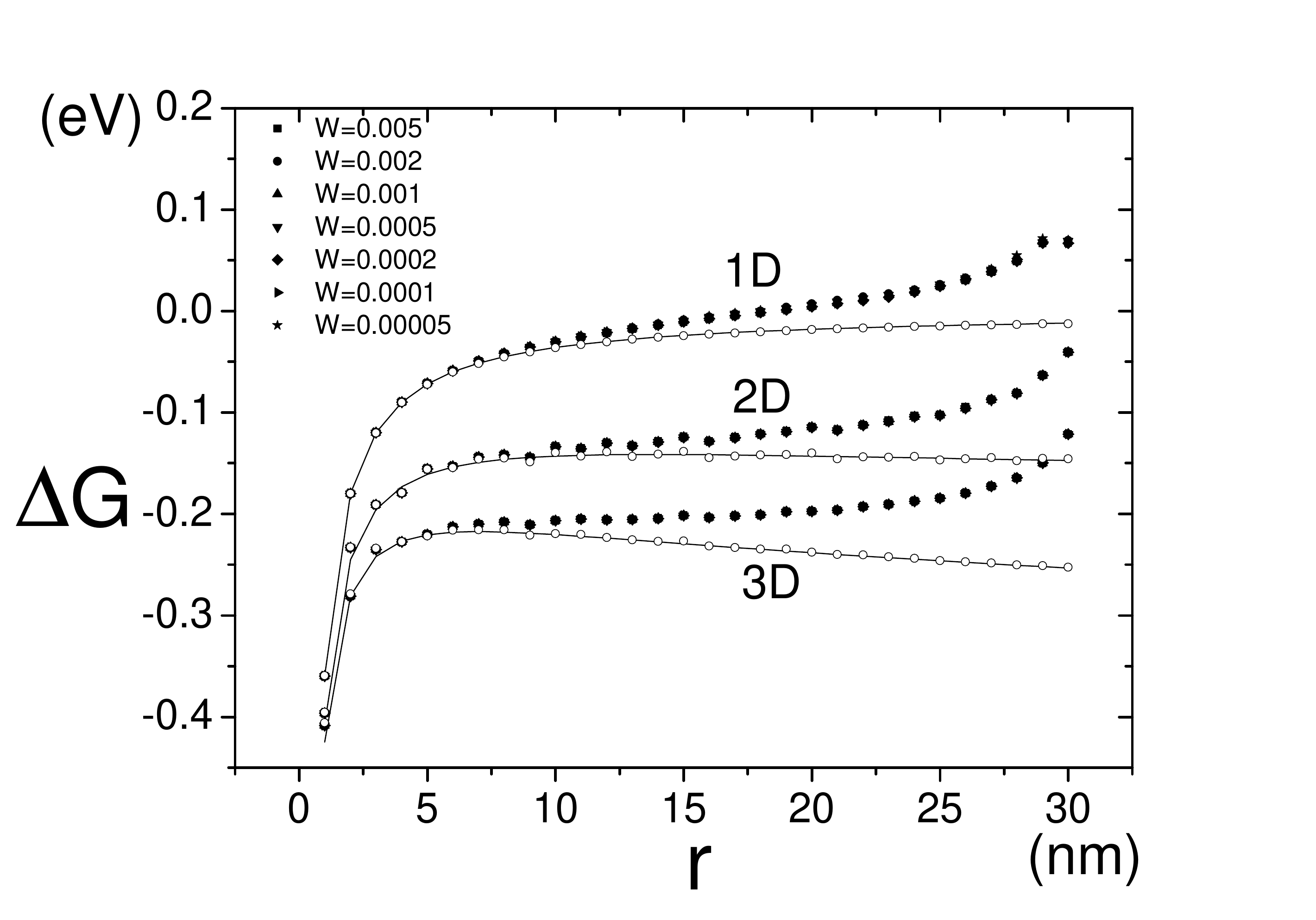}
\caption{The nonequilibrium free energy (full symbols) versus electron-hole distance for different recombination rates. Note that all full symbols contain seven (7) sets of data which all fall on the top of each other in each case (i.e the full symbols contain 21 sets of data).
Open circles are for the equilibrium Monte Carlo simulation results, the curves represent the analytical expression (\ref{eq9}) for $\Delta G$.}
\label{fig3}
\end{figure}

\begin{figure}[figure4]
\centering
\includegraphics[scale=.4]{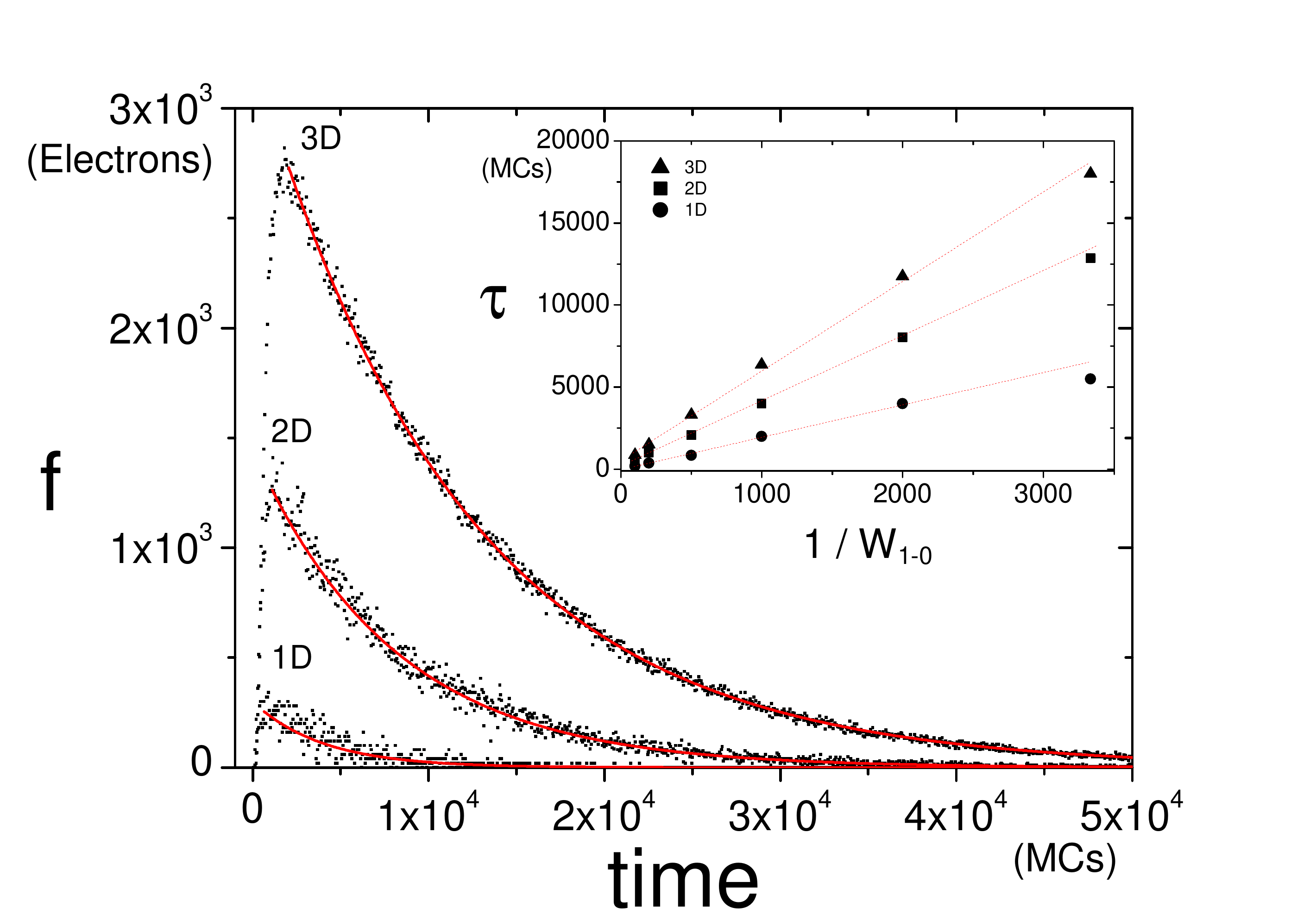}
\caption{(color online)The distribution of electron lifetimes before arriving at the collector for 1D, 2D and 3D systems for $w_{10}=0.0005$.
The frequencies for 1D and 2D systems are multiplied by 10 and 5, respectively. In the insert: the relaxation times of the arrival times distributions of the collected electrons versus the inverse recombination rates.} \label{fig4}
\end{figure}

\begin{figure}[figure5]
\centering
\includegraphics[scale=.4]{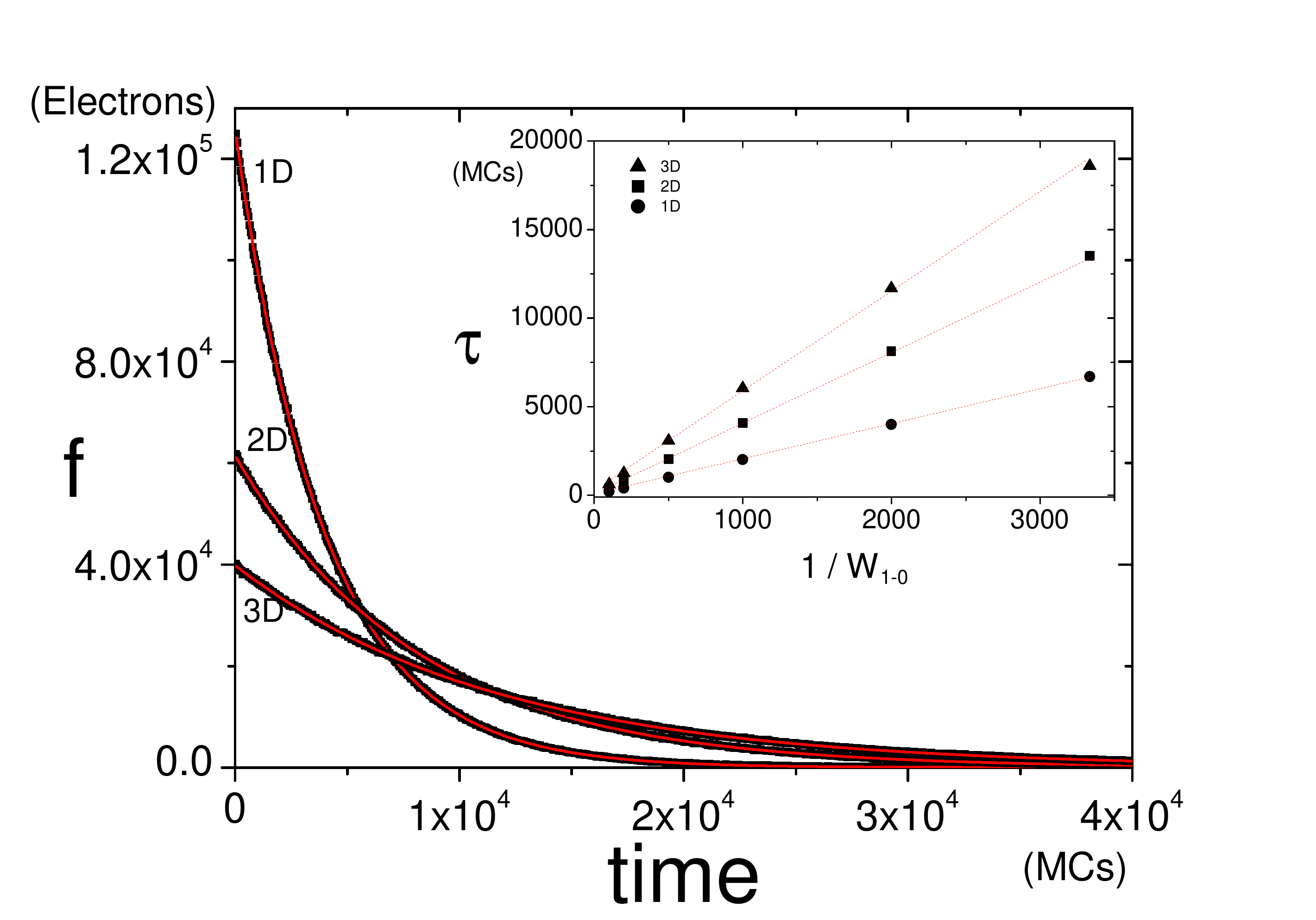}
\caption{(color online)The distribution of electron lifetimes before recombining for 1D, 2D and 3D systems for $w_{10}=0.0005$. The exponential
fits are given by the lines. In the insert: the relaxation times of the recombination times distributions versus the inverse recombination rates for the systems of different dimensionality; the lines are the results of linear fitting.} \label{fig5}
\end{figure}

Fig.\ref{fig4} demonstrates that the electron-hole separation process is strongly non-exponential during the initial period which is considerably
shorter than the subsequent exponential decay. Thus, although the criticism \cite{Wojcik2009} of the Onsager-Eigen-Braun type of models is reasonable,
the influence of this non-exponential character of the separation process can be of minor importance for recombination/dissociation, and these models can be used for analyzing experimental results.

\section{Conclusion}

Summarizing, the non-equilibrium simulation results show that the efficiency of charge separation in 3D systems is more than an order of
magnitude larger, as compared to 2D systems and almost three orders of magnitude larger than that in 1D systems at comparable conditions. The
ratio of efficiencies does not depend on the recombination rate if the electron yield is lower than 0.1. Surprisingly, the lifetimes of the
collected and recombined electrons are approximately equal and their values indicate that electrons perform long stochastic walks before they
are captured by the collector or recombined. This phenomenon is explained by the fact that only a small amount of electrons reaches the
collector and those ones that are able to move far from the origin can arrive at the collector or recombine with comparable probabilities. Just
these electrons determine the relaxation times. The simulation results compare well with the predictions of the continuum theory
\cite{Sano1979,Wojcik2009} that accounts for the finite recombination rate at nonzero reaction radius and non-exponential kinetics of the charge
separation process, if the reaction volume is taken equal to the total volume of primitive cells that are the nearest neighbors of the cell
where the hole is situated. This agreement validates the correspondence of the continuum and discrete models as well as the simulation results.

\label{}





\bibliography{bibliography}







\end{document}